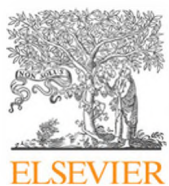
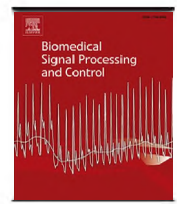

# Bi-LSTM neural network for EEG-based error detection in musicians' performance

Isaac Ariza [a,*], Lorenzo J. Tardón [a,*], Ana M. Barbancho [a], Irene De-Torres [b], Isabel Barbancho [a,*]

[a] *ATIC Research Group, ETSI Telecomunicación, Universidad de Málaga, 29071 Málaga, Spain*
[b] *Hospital Regional Universitario de Málaga, Av. de Carlos Haya, 84, 29010 Málaga, Spain*

ABSTRACT

Electroencephalography (EEG) is a tool that allows us to analyze brain activity with high temporal resolution. These measures, combined with deep learning and digital signal processing, are widely used in neurological disorder detection and emotion and mental activity recognition. In this paper, a new method for mental activity recognition is presented: instantaneous frequency, spectral entropy and Mel-frequency cepstral coefficients (MFCC) are used to classify EEG signals using bidirectional LSTM neural networks. It is shown that this method can be used for intra-subject or inter-subject analysis and has been applied to error detection in musician performance reaching compelling accuracy.

## 1. Introduction

Electroencephalography (EEG) is the measurement of signals coming from the electrical activity of the human brain; EEG signals allow real-time measurements of brain activity. The EEG signal is an oscillatory electrical potential generated by neuronal activations in the brain, recorded by electrodes placed on the scalp surface. The electrodes detect the small electrical signals that result from the activity of the brain cells. Only large populations of active neurons can generate enough electrical activity to be recorded on the scalp and even in such case, the amplitude of these signals is very small, in the μV range, due to their nature and the attenuation caused by the skull and tissues [1], so it is immediately amplified and then recorded, [2]. The study of the brain's activity through the EEG signals allows us to better understand how our brain work and which areas of the brain are activated when we carry out each type of activity.

Using digital signal processing techniques and machine learning methods, EEG signals can be analyzed to obtain useful results for different applications regarding neurological disorder detection [3–5], brain computer interfaces (BCI) [6,7], or emotion and mental task recognition [8–10]. Machine learning techniques are used in EEG signal analysis because they allow the extraction of features and classification using neural networks. In this sense, different kinds of neural networks have been used with this purpose, like convolutional neural networks (CNN) [11] and recursive neural networks (RNN), including long short term memory (LSTM) architectures, used to classify sequences [12]. LSTM neural networks have also been used for other purposes: Xiao et al. use dual-source LSTM networks to develop a model for underground coal gasification state prediction which attains an accuracy of 90.99% [13]; in [14], a convolution layer extracts spatial correlation and LSTM extracts temporal correlation in a Conv-LSTM network to perform predictions of a multivariate time series (MTS); other diverse specific architectures have been defined and employed [15] and, specifically, bidirectional LSTM (bi-LSTM) network architectures have been found useful in the speech processing framework [16], where context is important for the correct interpretation of the information.

In relation to the classification of EEG signals and music, machine learning techniques, SVM and MLP (Multilayer perceptron), are used to classify EEG signals into different emotions generated by listening to music with an accuracy of approximately 82% [17]; also LSTM architectures have been used in the EEG signal analysis context [18].

Previous works in relation with error detection during musician performance, analyze brain response when the performer commits an error [19]. Then, it was concluded that an error could be detected prior to its execution [20].

In this specific framework, it has been shown that playing piano may stimulate brain neuroplasticity. Plasticity describes changes in structure and function of the brain that affect behavior and that are related to experience or training [21]. Specifically, changes in the right auditory cortex have been seen in musicians [22], who, according to Pantev et al. [23], show enlarged auditory cortical evoked potentials to piano tones and this effect can be additionally modulated by the timbre of their own musical instrument.

The reward value and positive feedback associated with producing music might contribute to the possible modulation of neural plasticity via the reward circuitry [24], such positive or negative feedback are somehow related to our experiments.

In this paper, a new method for EEG signal classification for the detection of errors in the performance of piano and violin music based on Long-Short Term-Memory (LSTM) neural networks is presented; more specifically, the bidirectional LSTM architecture will be considered for its capacity to make use of both previous and later context, which the musician needs for the interpretation of the music performed or listened to. EEG signals coming from the recording of multiple EEG channels processed independently are employed. These signals, recorded during musicians' performance, are considered for classification into two categories: error event and correct event and, furthermore, the detection process is considered in both intra- and inter-subject scenarios.

The novelty of this method relies on the cross-analysis of the EEG signals, the intra- and inter-subject scenarios considered, the neural network architecture used and the features extracted for data characterization: Mel-frequency Cepstral Coefficients (MFCC) combined with other time–frequency domain features: instantaneous frequency and spectral entropy.

This paper is divided into four parts. First, the dataset employed and the data acquisition methodology are described. Then, the proposed method is exposed, including how trials are segmented in order to allow the extraction of features. The features themselves and the structure of the bi-LSTM neural network are later described. The results found after the experiments performed and a discussion on the results are exposed in Section 4. Finally, some conclusions are drawn in the last section.

## 2. Dataset employed

The description of the dataset used is provided next; this includes the depiction of the EEG acquisition system, the exposition of the audio recording system and the explanation of the dataset characteristics.

### 2.1. EEG Capture System

The BrainVision actiCHAMP-PLUS system is used to record EEG signals [25]. This system has a total of 128 channels, of which we use 64, positioned according to the 96Ch actiCAP snap standard.

Each channel is assigned a letter to identify the lobe: F (frontal), C (central), T (temporal), P (parietal) and O (occipital), and a number to define the position of the hemisphere: an odd number for the left side and an even number for the right side of the brain. The letter 'z' is used for the electrode used in the mid-line of the brain [2].

The reference channel is FCz and the ground channel is FPz. We move the electrodes of channels FT9 and FT10 near the eye to capture the vertical (VEOG) and horizontal (HEOG) ocular movements. We connect the electrode that correspond to Iz channel to a recording audio system. This audio recording is of low quality due to the system sampling rate, but it is used to allow the precise synchronization with the high quality audio recording performed as described in Section 2.2.

The channels used for audio and eye movements are, consequently, discarded regarding EEG analysis.

The electrodes used are active ones, which allow recordings to be made with low noise and high quality. The sample rate is 2500 Hz. The impedance of the electrodes is measured, making sure that their impedance is below 10 K in all cases. The impedance level is checked between each recording to ensure that it is kept as low as indicated.

The EEG signals obtained by this recording system have not been subjected to any type of pre-processing at the time of their acquisition.

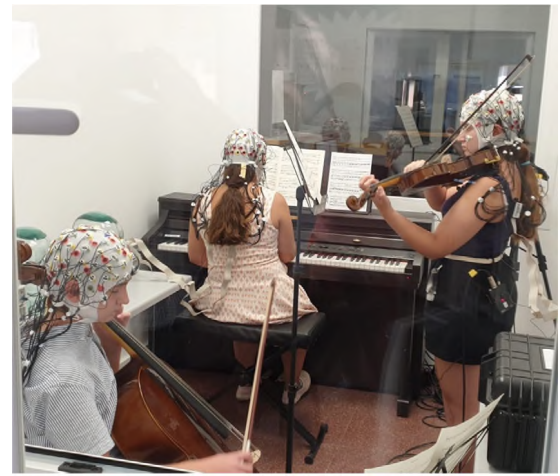

**Fig. 1.** Recording room and setup for audio and simultaneous EEG data acquisition.

### 2.2. Audio Capture System

A Zoom H6 have been used to audio record, configured with a sample rate of 48 kHz and using the maximum resolution available: 24 bits [26]. This device has 4 mono inputs and an aerial stereo microphone.

Three Zoom H6 mono audio inputs are used: L and R outputs of the Kawai CA91 piano are connected to two of them and an AKG C-411 contact microphone placed on the violin, which picks up the vibration of the violin avoiding the sound of the piano, is connected to the third mono input used. In this way, we get separate audio from the violin and the piano. Also, the Zoom's stereo microphone is used to obtain the recording of the aerial sound to obtain the mix of the piano and the violin.

A 27'' iMac with Logic Pro X connected to the MIDI output of the piano Kawai CA91 is employed. Thus, the MIDI output of the piano are recorded at the time. This MIDI recording makes it easier for us to detect errors in the piano performance.

Finally, in order to synchronize EEG and audio recordings, a StimTrack device is used. A clap is recorded in the low quality audio acquired with the selected EEG channel as well as in the high quality aerial recording. Also, a trigger is launched by StimTrack to facilitate the synchronization process.

### 2.3. Dataset characteristics

The dataset used consists of the recording of the audio and EEG signals of a pianist and a violinist during the performance of a piano trio composition. The data acquisition methodology was conducted according to the guidelines of the Declaration of Helsinki, and approved by the Comité de Ética de la Investigación Provincial de Málaga on December 19, 2019, approval number: 2176-N-19. Informed consent was obtained from all subjects involved in the study.

Specifically, recordings were done during the performance of a pianist, a violinist and a cellist of the Joseph Haydn, Piano Trio in G Major, Hob. XV:25, "Gypsy Trio": 1. Andante. This first movement lasts around 4 min and a total of 5 recordings were made. The performance of the cellist was also recorded, however it has not been used because the number of samples available was too small. Fig. 1 shows the recording setup

As it has just been anticipated, in order to label the EEG samples, it is necessary to label the audio recordings. To do this, the recordings are listened to and compared with the musical score to find the mistakes in the notes played by the interpreters. Five different labels are used. These labels, together with their meaning are shown in Table 1. We consider NCH, OOT and MIS as error events (ERR). For each event, we store the starting time, the ending time and the event type.

**Table 1**
Dataset: Labels employed for audio annotation.

| Label | Meaning |
| --- | --- |
| OK | Correct performance |
| NCH | Note changed with respect to score |
| OOT | Out of tone |
| SIL | Silence not indicated in the score |
| MIS | Missing note with respect to score |

**Table 2**
Summary of labeled piano excerpts.

| Label | Number of excerpts | Total duration (s) | Number of trials |
| --- | --- | --- | --- |
| OK | 40 | 719.82 | 597 |
| NCH | 36 | 74.90 | 55 |
| SIL | – | – | – |
| MIS | 6 | 13.90 | 8 |

**Table 3**
Summary of labeled violin excerpts.

| Label | Number of excerpts | Total duration (s) | Number of trials |
| --- | --- | --- | --- |
| OK | 16 | 367.43 | 301 |
| NCH | 4 | 7.15 | 6 |
| OOT | 9 | 32.56 | 28 |
| SIL | 2 | 6.44 | 5 |
| MIS | – | – | – |

Tables 2 and 3 show the number of excerpts, duration and the number of trials of each type of event for the violin and piano recordings.

The audio samples are divided into trials of 1 second of duration. Then, each 1-second trial is assigned the corresponding label according the analysis of the audio recordings; trials that do not completely lay inside a certain labeled excerpt are discarded to avoid ambiguity. With all this, 597 correct-performance trials (OK) and 63 error trials for the piano and 301 correct-performance trials (OK) and 34 error trials for the violin are available.

Once the audio is tagged, it is necessary to synchronize audio and EEG recordings. This task is carried out by looking for the first peak of the envelope, which corresponds to a clap used as marker for the start of the recording.

Note that although the dataset is small, it serves to extract conclusions on the feasibility and work of music performance error detection using EEG in the scenario considered. Also, note that there are no other datasets of synchronized labeled (from the musical point of view) EEG and music signals like the one employed in this study, to the authors knowledge.

## 3. Analysis methodology

In this section, we describe the specific way in which the data are prepared and processed and the classification scheme devised. Next, the process of data preparation, labeling and selection of train and test subsets are explained. Then, in Section 3.2, the selected features to obtain from the signal extracts to characterize the EEG signals are exposed. Finally, the architecture of the bidirectional LSTM neural network used to classify the EEG signals is shown.

### 3.1. Data preparation

The EEG data are split into 1 s length trials. Thus, each EEG trial is related to a data matrix with 61 rows, one for each channel.

In order to process the EEG signal, each channel is treated independently. Fig. 2 shows the process to segment and label EEG signals.

Each EEG trial is individually identified in a trial by trial basis; thus each 1 s trial is assigned a label to identify the two states considered:

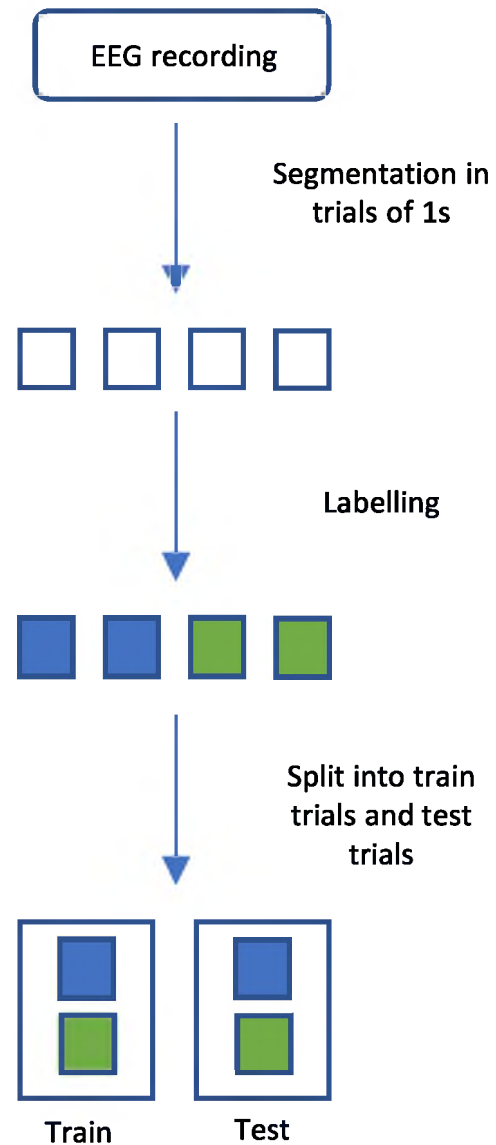

**Fig. 2.** Illustrative diagram of the process followed to segment, label and split trials into train and test subsets.

correct performance, error in performance. After the samples are labeled, they are split randomly into two sets of trials, a set to train the neural network and another one to test the performance.

### 3.2. Signal characterization

In order to classify the EEG trials, three different features are extracted for each channel: the instantaneous frequency, the spectral entropy and the Mel-frequency cepstral coefficients (MFCC).

Each trial is processed on a channel by channel basis. This means that for each of them, the instantaneous frequency, the spectral entropy and 11 MFCCs are calculated, consequently, an N×13 matrix is created to characterize each trial. Recall that, in this studio, N=61 due to the fact that we use 61 EEG channels.

This process is depicted in Fig. 3. Each row of the matrix drawn is a vector input to the bi-LSTM neural network.

Further details about the features employed to characterize the EEG signals are given next.

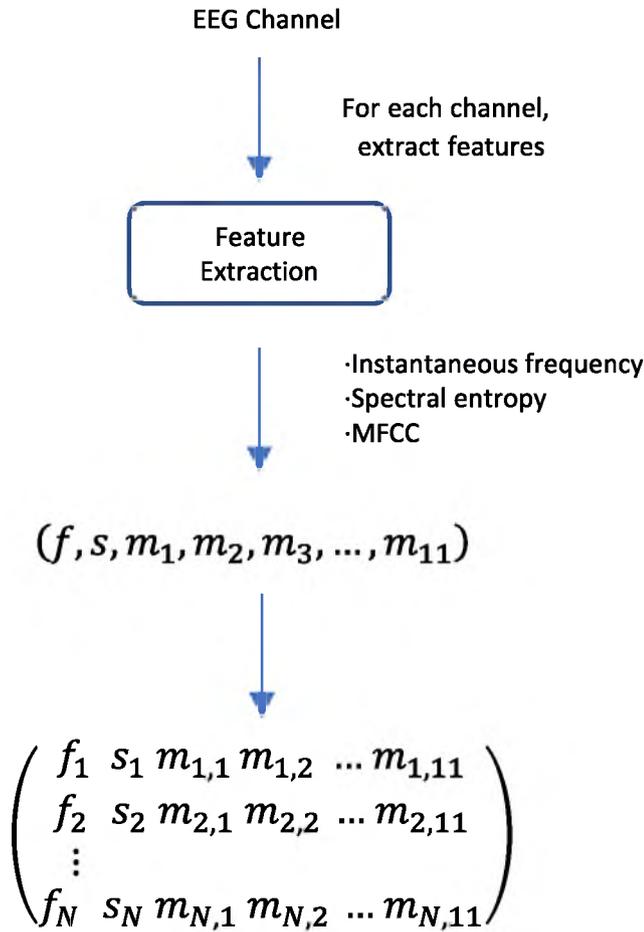

Fig. 3. Illustration of the construction of the N×13 array that represents each 1 s trial.

### 3.2.1. Instantaneous frequency

The instantaneous frequency for non-stationary signals is a time-varying parameter that defines the location of the main spectral peak of the signal as it changes over time [27]. Conceptually, it can be interpreted as the frequency of a sinusoidal wave locally matched to the signal under analysis.

In order to calculate this feature, Matlab's *instfreq* function [28] is employed with the default parameters. This function estimates the instantaneous frequency as the first conditional spectral moment of the time–frequency distribution of the input signal as follows:

$$f_{inst}(t) = \frac{\int_0^\infty f P(t,f) df}{\int_0^\infty P(t,f) df} \quad (1)$$

where $P(t,f)$ is the power spectrum of the input signal at frame $t$.

### 3.2.2. Spectral entropy

Entropy, in information theory, describes the average information of a random event or a random variable. Mathematically, this parameter is defined as [29]:

$$H = -\sum_{i=1}^{N} p(a_i) \log_2 p(a_i) \quad (2)$$

where $p(a_i)$ stands for the probability of the discrete random variable $A$ taking the value $a_i$.

Specifically, we make use of the spectral version of the entropy: the instantaneous spectral entropy [30]. This feature can be easily calculated using Matlab's function *pentropy* [31] that, conveniently, uses the normalized time–frequency power spectrogram $P(t,m)$ as follows:

$$H(t) = -\sum_{m=1}^{N} P(t,m) \log_2 P(t,m) \quad (3)$$

where $P(t,m)$ is obtained using the power spectrogram $S(t,m)$ as follows:

$$P(t,m) = \frac{S(t,m)}{\sum_{k=1}^{N_F} S(t,k)} \quad (4)$$

with $t$ indexing the time frame and $N_F$ representing the number of samples of the DFT.

### 3.2.3. Mel-frequency cepstral coefficients

One of the most efficient tools for audio and music characterization are the Mel frequency cepstral coefficients [32,33]. These are defined upon the critical bandwidth which is the bandwidth around a central frequency that the human hearing system needs to distinguish a different tone. This bandwidth can be defined using this expression:

$$BW_{critical} = 25 + 75 \left[1 + 1.4 \left(\frac{f}{1000}\right)^2\right]^{0.69} \quad (5)$$

Mel filters take these critical bands into account to define filter banks in which the individual filters' bandwidth grows logarithmically. Also, each filter starts at the location of the peak of the previous filter.

Specifically, the following conversion is employed to convert from Hertz to Mel units [34]:

$$f_{MEL} = 2595 \log_{10}\left(1 + \frac{f(Hz)}{700}\right) \quad (6)$$

which is employed to create the filters evenly spaced in MEL units. Note that their design is adapted to the EEG sampling rate, which is 2.500 in our recordings, and the frequency limit specified, which in this case is 200 Hz. This limit is large enough to gather all the necessary information from the EEG signals for our task, at the sight of the commonly used frequency bands [35].

Now, in order to calculate the Mel frequency cepstral coefficients, the following steps are followed: divide the signal into 45 ms segments with an overlap of 10 ms, obtain the DFT of each segment and estimate the periodogram of the signal frame:

$$P_i(k) = \frac{1}{N}|X_i(k)|^2 \quad (7)$$

where $|X_i(k)|$ represent the module of the DFT of segment $i$ of a certain frame.

On the other hand, the Mel filters are obtained by applying the following expressions:

$$h_m(k) = \begin{cases} 0 & k \leq f(m-1) \\ \frac{k-f(m-1)}{f(m)-f(m-1)} & f(m-1) \leq k \leq f(m) \\ \frac{f(m+1)-k}{f(m+1)-f(m)} & f(m) \leq k \leq f(m+1) \\ 0 & k > f(m+1) \end{cases} \quad (8)$$

In order to employ these equations, the centers and limits of the filters are rounded to coincide with the DFT samples (Fig. 4).

After the MEL filters are properly applied, the process to obtain the MFCC continues following the conventional MFCC stages [33].

### 3.3. Bidirectional LSTM Neural Network (bi-LSTM)

A type of a Long Short Term Memory (LSTM) neural network is used for EEG signal classification [36]. This kind of neural network, in which its nodes control the flow of information, is specially used for the classification of time sequences since they make use of long-term dependencies in the signal. The following steps are performed in the network [37]:

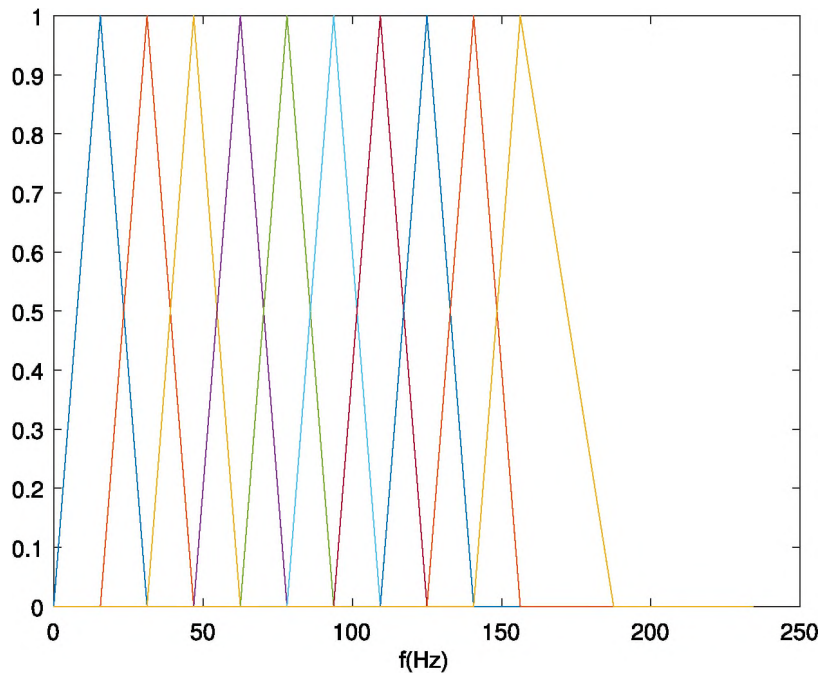

**Fig. 4.** Adapted Mel filters employed to obtain MFCC for EEG signal characterization.

1. Forget the irrelevant information.
2. Perform the computations to store relevant parts of new information.
3. Use the two previous steps to selectively update its internal state.
4. Generate an output.

Specifically, it is this temporal characteristic that made us to choose this type of network for our task. However, similarly to the speech recognition framework, the correct interpretation of music, played or listened to, requires some previous and posterior context, which makes the bidirectional-LSTM architecture perfectly adapted to our case. This variety of LSTM architecture contains layers that can be trained in both forward and backward time directions simultaneously [15], which provides a signal analysis context more similar to the one a musician employs to interpret music.

The neural network employed is composed of an input sequence layer of $N$ inputs, with $N$ the number of EEG channels. Each input a vector of 13 elements (Fig. 3) built upon the instantaneous frequency, the spectral entropy and the MFCCs, which were described earlier. These correspond to every 1 s trial available.

Then, a bidirectional LSTM layer, with 20 hidden units, that learns bidirectional long-term dependencies between sequence data follows. The number of hidden units indicates the amount of information remembered between time steps. Note that in addition to the bi-LSTM architecture that contains a bi-LSTM layer, as shown in Fig. 5, we have tested the network architecture with a conventional LSTM layer and a Gated Recurrent Unit (GRU) layer [15].

A fully-connected layer that determines the number of outputs is then added. In this study, the number of outputs is 2: error trial and correct trial.

Finally, a non-linear softmax layer and a classification layer that computes the cross entropy loss complete the network architecture. An illustrative scheme of the network architecture is drawn in Fig. 5.

## 4. Results

In this section, the results obtained for the different kinds of experiments carried out are shown. We have performed evaluations for intra-subject and inter-subject scenarios using the data available.

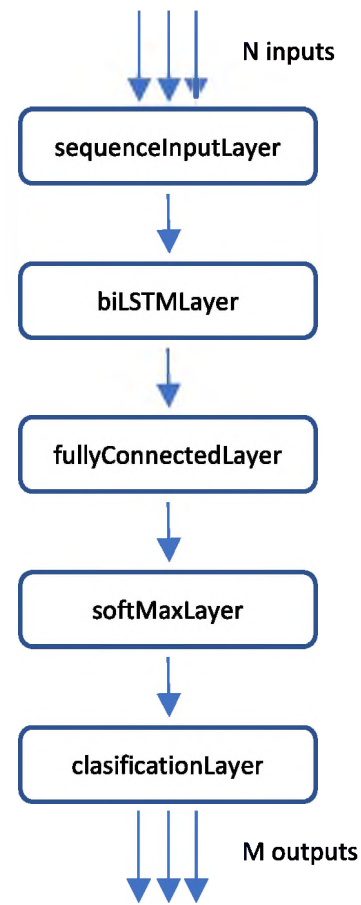

**Fig. 5.** Illustration of the architecture of the bi-LSTM neural network designed to classify EEG signals.

First, data regarding the recording of a pianist playing the Joseph Haydn, Piano Trio in G Major, Hob. XV:25, "Gypsy Trio": 1. Andante

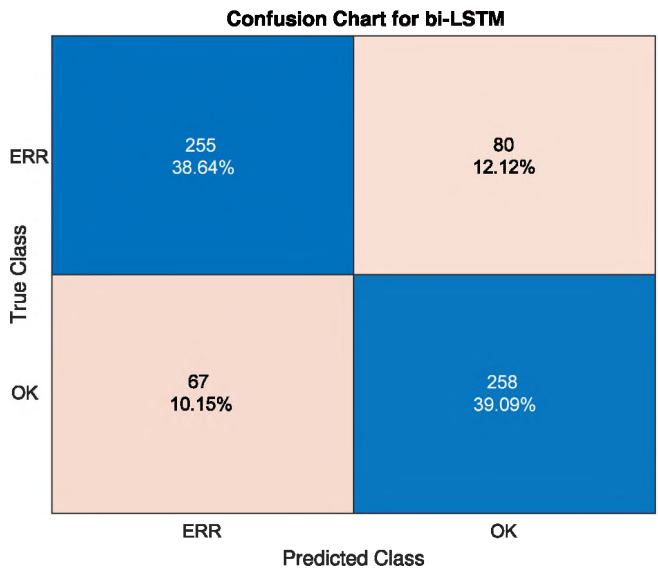

Fig. 6. Confusion chart for the intra-subject error detection evaluation test using EEG data of the pianist performance.

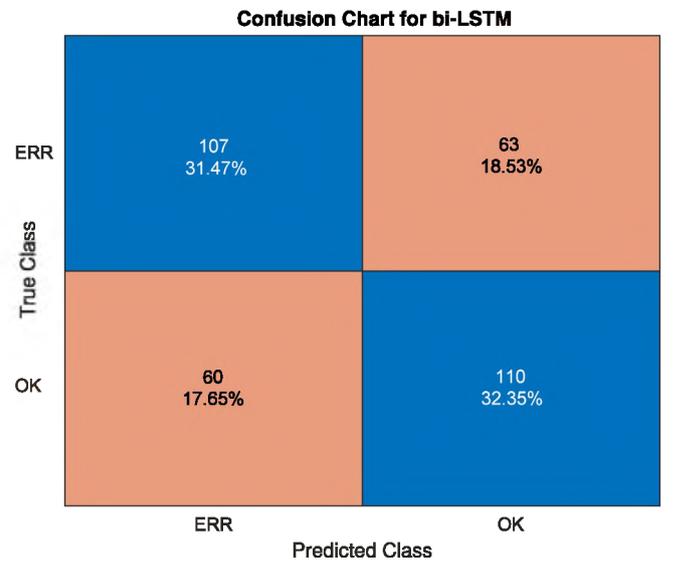

Fig. 7. Confusion chart for the intra-subject error detection evaluation test using EEG data of the violinist performance.

are used. These data have been used for both training, validating and testing the neural network. These tests are intra-subject since the data employed comes from the same subject in both training and evaluation tasks.

This intra-subject scheme is then repeated for the recording of the violinist playing the same piece.

Later, the inter-subject scenario is evaluated. In this case, the pianist data is considered, in the first scenario, to train a neural network and, then, that trained neural network is employed to classify trials defined on the data of the violinist playing the same musical piece. In the second scenario, the roles of the piano and violin data are swapped.

Note that for all the tests, the bi-LSTM layer of the neural network is configured with 20 hidden units and trained the network for 5 epochs. The same configuration is employed for the two other network architectures, LSTM and GRU, considered for comparison.

In the following subsections, the results obtained and some comments on them are drawn.

*4.1. Intra-subject scenarios*

The performance of the proposed scheme for intra-subject performance detection is obtained using cross-validation [38]. Specifically, trials are split into two different subsets: one to train the neural network and another one to measure the performance. The selection of training samples is done randomly using Matlab's *dividerand* function [39]. Note that the size of the training set is significantly larger than the test set, thus under sampling is considered and the evaluation is repeated 10 times for each channel (10-fold cross-validation).

First, 660 trials of a pianist are used; among them, 63 correspond to error events (see Section 2.3) and the rest correspond to correct performance.

Unbalanced classes are a drawback for classification so under sampling is used to have the same number of correct and wrong samples: 63 trials of both correct and wrong performance.

Fig. 6 shows the confusion chart of the detection of error trials using the EEG data processed and characterized as described in Section 3.2 and the bi-LSTM network described in Section 3.3.

Regarding F-measures, the accuracy attained is 75.76%, F-score is 78.13%, recall is 75.76% and precision is 80.65%. It should be noted that no distinction has been made between the different types of error since the number of samples per type is not large.

The accuracy attained by the other network types considered for comparison, under the same conditions, is 61.12% and 54.55% for the LSTM and GRU architectures, respectively.

Then, the EEG data corresponding to the synchronous performance of the violin on the Haydn trio are considered. A total of 335 samples among which 34 correspond to error trials are available. Again, under sampling is employed to evaluate the error detection ability using the EEG data of the violinist for both training and testing. Again, 10-fold cross-validation, randomizing the selection of the available samples is performed. Fig. 7 shows the confusion chart of this second scenario evaluated.

In this experiment, the accuracy obtained is 61.76%, F-score is 62.50%, recall is 58.82% and precision is 62.50%. If these intra-subject results are compared with those obtained with the piano, it must be observed that the performance is worse. We deem this is due to the greater variety of types of errors in the violin, together with the greater subjectivity in the perception of these errors. Examples of this issue are the out-of-tune note error event which, in some cases, are difficult to perceive even in the off-line labeling process.

The accuracy attained by the LSTM and GRU architectures in this experiment is 60.58% and 60.59%, respectively.

*4.2. Inter-subject scenarios*

After the intra-subject tests, we decided to consider the inter-subject detection scenario. In this specific situation, two different people play two different instruments: a piano and a violin.

Note that the variability of the data is very high now. This scenario is considering different subjects, different instruments and even diverse types of errors in the performance. It must be noted that most of the violin's errors identified are out of tone errors (OOT); however, this type of error does not exist on the piano.

The bi-LSTM neural network is trained with the data of the pianist. Concretely, the network is trained with 132 samples; of which 66 correspond to error events of 1 s of duration and the rest correspond to trials where no error is made. The evaluation of the detection behavior using the EEG is done using the violinist data: 68 trials are used, with 34 corresponding to error. In these conditions, the accuracy obtained is 55.88%, F-score is 64.52%, recall is 58.82% and precision is 62.50%; Fig. 8 draws the confusion chart for this experiment. The LSTM and GRU architectures attain lower performance: 55.29% and 52.94% of accuracy, respectively.

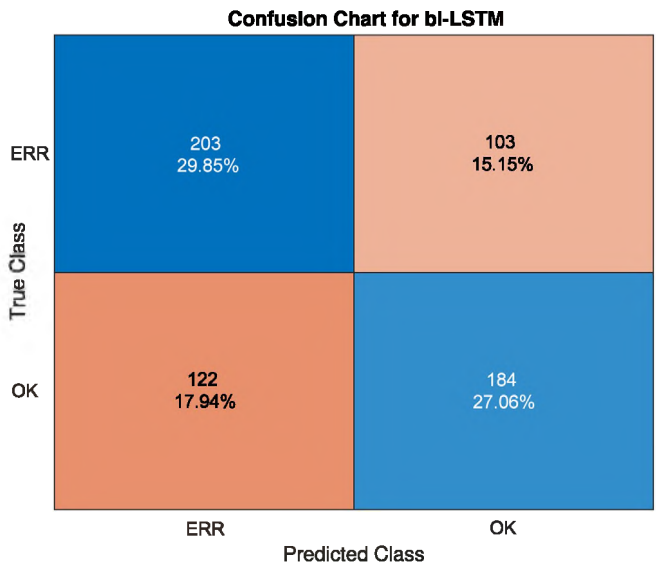

**Fig. 8.** Confusion chart for the first inter-subject test (training is performed with the pianist data and testing is carried out using the violinist data). The chart shows the total number of true and false positive and negative detection decisions.

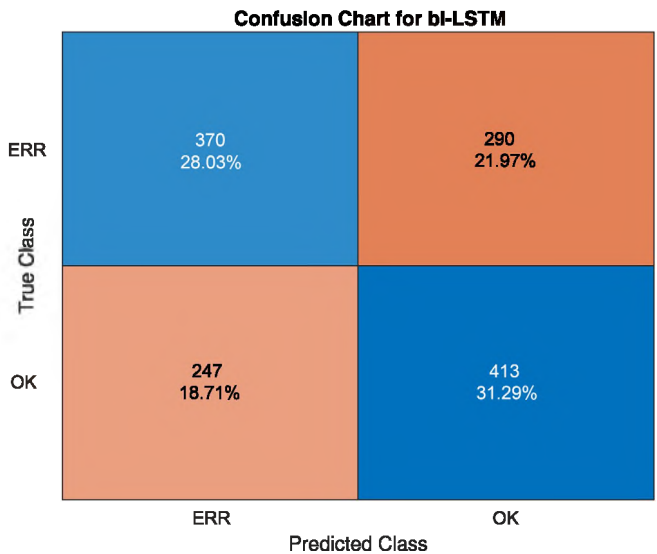

**Fig. 9.** Confusion chart for the second inter-subject test (training is performed with the violinist data and testing is carried out using the pianist data). The chart shows the total number of true and false positive and negative detection decisions.

Another similar evaluation step is performed switching the roles of piano and violin data; e.g. training is carried out with violin data and the performance is evaluated with piano data. The accuracy attained in this case is 59.09%, F-score is 58.27%, recall is 56.06% and precision is 60.66%. In Fig. 9, the corresponding confusion chart is shown. Note that these results do not largely differ from the previous inter-subject ones nor from the violinist intra-subject test. The LSTM and GRU architectures attain an accuracy 55.91% and 48.94%, respectively.

### 4.3. Discussion

Remarkably, in spite of the differences in the results, the performance of the bi-LSTM network with the features selected seems to confirm the validity or the features and classification strategy for the detection of errors in the performance of both piano and violin.

In the intra-subject scenario, two tests have been carried out: a test with data of a pianist and other with data of a violinist. The highest F-score attained was 78.13% with the pianist data. The results of the second intra-subject test are worst than in the previous one, attaining 62.5% of F-score. This is due to the smaller number of violin trials, which has an impact on the training of the network, but also on the larger variability or the error events in the violin performance and the consequent larger ambiguity and difficulty for its perception and even its annotation.

Regarding the inter-subject case, the number and range of the diverse variables involved in the experiments is even larger than in the previous cases: we use EEG signals from different people and, in addition, different instruments are played. Additionally, It should also be noted that most of the violin errors are out of tone errors; this king of error does not exist in the piano, which can explain the rise of the difficulty for the performance in the inter-subject case. Also, we have significantly fewer trials for the violin.

Nevertheless, it was found remarkable the fact that, to a certain extent, using the features selected and the defined bi-LSTM network it is possible to perform our detection task under such conditions. This can be justified on the basis of similitudes in the perception of musical errors in the performance of the players created upon their musical training and brain plasticity.

Also recall that the performance attained by the other architectures evaluated: LSTM and GRU, is lower, which can be significantly due to the lack of context derived from the different architecture with respect to the selected bi-LSTM network.

## 5. Conclusion

In this paper, the possibility of detection of errors in the performance of music players using recorded EEG signals, characterized by their instantaneous frequency, spectral entropy and 11 Mel-frequency cepstral coefficients and processed by a bidirectional LSTM neural network has been investigated.

Specifically, this scheme has been used to detect errors in the performance of musicians playing a piano and a violin using their respective EEG signals; intra- and inter- subject scenarios have been addressed. In the former, the neural network is trained with data from the same musician that will be considered for the evaluation of the classification performance; in the inter-subject scenario, training data comes from a certain musician whereas test data comes from the other musician playing a different instrument. The devised scheme attained significant performance in the musical error detection task considered.

The proposed method attained F-score over 78% in the intra-subject case corresponding to the pianist, whereas the violin only scenario attained over 62% of F-score. This reduction of the performance is considered to be caused by the lower number of samples for training and the larger variability in the error characteristics and their perception.

The intra-subject scenario, which implies different subjects playing diverse instruments for training and evaluation attains an average F-score of 61.4%; the ability to perform musical error detection using EEG data in such scenario, to a certain extent, is explained, from our point of view, by the similitude in the perception of music and training of the brain in the musical context; on the other hand, the descent in the performance score is easily explained by the large variability in the evaluation scenario.

Further improvements and experiments could use, at the beginning of the signal processing stage, independent component analysis (ICA) or the identification of reduced number of EEG channels or brain areas to analyze. Also, the scheme and analyses would benefit from the acquisition of additional recordings.

**CRediT authorship contribution statement**

**Isaac Ariza:** Conceptualization, Methodology, Software, Investigation, Writing. **Lorenzo J. Tardón:** Conceptualization, Methodology, Software, Investigation, Writing, Project administration. **Ana M. Barbancho:** Conceptualization, Methodology, Software, Investigation, Validation, Writing. **Irene De-Torres:** Conceptualization, Methodology, , Investigation, Validation, Writing. **Isabel Barbancho:** Conceptualization, Methodology, Software, Investigation, Writing, Validation, Project administration.

**Declaration of competing interest**

The authors declare that they have no known competing financial interests or personal relationships that could have appeared to influence the work reported in this paper.

**Acknowledgments**

This research was possible thanks to the recording of the Joseph Haydn, Piano Trio in G Major, Hob. XV: 25, 'Gypsy Trio': I. Andante performed by Isabel Barbancho, Isabel Tardón and Jorge Tardón, who played the piano, violin and cello, respectively, while recording their EEG signals. The authors are also grateful to Samuel Villalba for labeling the audio samples employed.

This work has been funded by Junta de Andalucía in the framework of Proyectos I+D+I en el marco del Programa Operativo FEDER Andalucia 2014–2020 under Project No.: UMA18-FEDERJA-023, Proyectos de I+D+i en el ámbito del Plan Andaluz de Investigación, Desarrollo e Innovación (PAIDI 2020) under Project No.: PY20_00237 and Universidad de Málaga, Campus de Excelencia Internacional Andalucia Tech. Funding for open access charge: Universidad de Málaga/CBUA.